\documentclass[nofootinbib,prd,onecolumn]{revtex4}
\usepackage{amsmath}
\usepackage{graphicx}
\usepackage{amssymb}
\usepackage{float}
\usepackage{mathpazo}
\usepackage{lineno}
\usepackage{hyperref}

\pdfoutput=1
\newcommand{\ignore}[1]{}

\newcommand{\be}{\begin{equation}}
\newcommand{\ee}{\end{equation}}
\def \ba#1\ea{\begin{align}#1\end{align}}
\newcommand{\bit}{\begin{itemize}}
\newcommand{\eit}{\end{itemize}}

\newcommand{\e}{\mathrm{e}}

\def \slashb#1{\setbox0=\hbox{$#1$}#1\hskip-\wd0\dimen0=5pt\advance
        \dimen0 by-\ht0\advance \dimen0 by\dp0\lower0.5\dimen0\hbox
          to\wd0{\hss \sl/\/ \hss}}

\input{epsf}
\begin{document}

\title{New physics from COHERENT data with improved Quenching Factor}
\author{Amir N.\ Khan$^{a, b}$}
\email{amir.khan@mpi-hd.mpg.de, akhan@fnal.gov}
\author{Werner Rodejohann$^{a}$}
\email{werner.rodejohann@mi-hd.mpg.de}
\affiliation{$^{a}$Max-Planck-Institut f\"ur Kernphysik, Postfach 103980, D-69029
Heidelberg, Germany\\
$^{b}$Theoretical Physics Department, Fermi National Accelerator Laboratory,
P.O. Box 500, Batavia, IL 60510, USA}

\begin{abstract}
\noindent
A recent new measurement and re-analysis of past measurements suggested an improved quenching factor value and  uncertainty for CsI[Na]. This implies a measurement of the COHERENT experiment of coherent elastic neutrino-nucleus scattering that is closer to the Standard Model prediction and has less uncertainty. 
We illustrate the impact of this improvement by revisiting fits to the Weinberg angle, neutrino magnetic moments, neutron rms and neutrino charge radii, weak nuclear charge of the Cs nucleus, neutrino non-standard interactions (in particular those relevant for LMA-Dark) and new scalar as well as vector bosons. Significant improvement is observed, particularly for those scenarios coherently affecting the electroweak SM process.

\end{abstract}

\pacs{xxxxx}
\maketitle

\section{Introduction}
\noindent
Coherent elastic neutrino-nucleus scattering (CE$\nu $NS) has been predicted in 1974 \cite{Freedman:1973yd}, but not been observed before 2017 \cite{Akimov:2017ade}. The importance of the process ranges from its ability to probe Standard Model parameters at low momentum transfer \cite{Scholberg:2005qs,Lindner:2016wff,Miranda:2019wdy}, test new neutrino physics and new neutral currents in general \cite{Barranco:2005yy, Dutta:2015vwa,Lindner:2016wff, Dent:2016wcr, Coloma:2017ncl, Liao:2017uzy, Dent:2017mpr, Kosmas:2017tsq, Farzan:2018gtr, Abdullah:2018ykz, Bauer:2018onh,Heeck:2018nzc, Denton:2018xmq, Billard:2018jnl, Altmannshofer:2018xyo, AristizabalSierra:2019zmy,  Miranda:2019skf, Dutta:2019eml,  AristizabalSierra:2019ufd, Bischer:2019ttk,Arcadi:2019uif}, sterile neutrino searches \cite{Anderson:2012pn, Dutta:2015nlo, Kosmas:2017tsq, AristizabalSierra:2019zmy, Miranda:2019skf}, implications for supernova physics \cite{Freedman:1977xn,Melson:2019kjj,Raj:2019wpy,Raj:2019sci}, dark matter searches \cite{deNiverville:2015mwa,Ge:2017mcq,Brdar:2018qqj, Dutta:2019nbn,Chao:2019pyh}, neutrino magnetic moments \cite{Dodd:1991ni,Scholberg:2005qs, Kosmas:2015sqa, Kosmas:2017tsq, Billard:2018jnl, Miranda:2019wdy}, nuclear physics \cite{Cadeddu:2017etk,Huang:2019ene,Papoulias:2019lfi,Ciuffoli:2018qem} and its connection to gravitational waves \cite{Hagen:2015yea,Wei:2019mdj}. The process under discussion is taking place at energies below about 50 MeV, and given by 
\begin{equation}
\nu + N \to \nu + N \, .
\end{equation}
Nuclear recoil is the relevant observable. 
In case of COHERENT, a CsI[Na] scintillation detector was used as target. 
Experimentally, the so-called Quenching Factor (QF) is of crucial importance. It denotes the energy-dependent ratio of the scintillation signal from nuclear recoils with respect to the one from electron recoils, i.e.\ the ratio of recorded energy to nuclear recoil. In the publication of the COHERENT experiment the QF-uncertainty of 18.9\% dominated the total uncertainty \cite{Akimov:2017ade}. Recently, past measurements of the QF were revisited, and a new one was performed \cite{Collar:2019ihs}. As a result, new (energy-dependent) values for the QF and its uncertainty were proposed. 
Applied to COHERENT, the systematic uncertainty would reduce from 28\% to 13.5\%, and the SM-predicted  rate would reduce from 
$173 \pm 48$ to $138 \pm 19$, compared to the measurement of $134 \pm 22$ events. 
As suggested in \cite{Collar:2019ihs}, physics extracted from the measurement would 
significantly improve when taking into account the new QF-values and uncertainty. 

We perform in this paper, as illustration of the impact of improved quenching understanding, a fit to COHERENT data taking into account the new QF value and uncertainty. We consider several parameters that can be extracted from CE$\nu$NS, namely the Weinberg angle, neutrino magnetic moment and charge radii, the neutron rms charge radius, neutrino non-standard interactions, as well as couplings and masses of new vector and scalar bosons mediating CE$\nu$NS. Improvement is found, particularly for those scenarios that coherently affect the SM process.

The paper is built up as follows. In Section \ref{sec:data} we describe the data we fit and the procedure we follow. Section \ref{sec:EW} gives the fit results for parameters related to SM and new electroweak physics, namely the Weinberg angle, neutrino magnetic moments and neutron/neutrino charge radii. 
Section \ref{sec:NSI} deals with neutrino non-standard interactions, Section \ref{sec:bosons} with parameters related to new vector or scalar bosons, before we conclude in Section \ref{sec:concl}.

\section{\label{sec:data}COHERENT Data and Fit Procedure}
\noindent 
The neutrino source for COHERENT's detection of coherent elastic neutrino-nucleus scattering are pions  produced from the spallation neutron source. The total number of protons on target (POT) delivered to a liquid mercury target was  $N_{\mathrm{POT}}^{\rm tot} =1.76\times 10^{23}$ \cite{Akimov:2017ade}.  Monoenergetic muon neutrinos $%
(\nu _{\mu })$ at $E_{\nu }=29.9$~MeV are produced from pion decay at rest ($\pi ^{+}\rightarrow \mu ^{+}\nu _{\mu })$, followed by a delayed
beam of electron neutrinos ($\nu _{e})$ and muon-antineutrinos ($\bar{\nu}%
_{\mu })$ produced subsequently by muon decay $\mu ^{+}\rightarrow \nu
_{e} \, e^{+}\bar{\nu}_{\mu }$. The 
average production rate from the pion decay chain is $r=0.08$ neutrinos of each flavor per proton. 

The CsI[Na] scintillator 
detector is located at a distance of $L = 19.3$ m. The fluxes are \cite{Akimov:2018vzs} 
\begin{eqnarray}
\frac{d\phi _{\nu_\mu }(E_{\nu })}{dE_{\nu }} &=&\frac{rN_{pot}}{4\pi L^{2}}%
\delta \left(E_{\nu }-\frac{m_{\pi }^{2}-m_{\mu }^{2}}{2m_{\pi }}\right) ,  \notag \\
\frac{d\phi _{\overline{\nu }_\mu }(E_{\nu })}{dE_{\nu }} &=&\frac{rN_{pot}}{%
4\pi L^{2}}\frac{64E_{\nu }^{2}}{m_{\mu }^{3}}\left(E_{\nu }-\frac{m_{\pi
}^{2}-m_{\mu }^{2}}{2m_{\pi }}\right),  \notag \\
\frac{d\phi _{\nu_e}(E_{\nu })}{dE_{\nu }} &=&\frac{rN_{pot}}{4\pi L^{2}}%
\frac{192E_{\nu }^{2}}{m_{\mu }^{3}}\left(\frac{1}{2}-\frac{E_{\nu }}{m_{\mu }}\right), 
\end{eqnarray}%
where $N_{pot}=5.71\times 10^{20}$ are the number of protons per day.  The differential cross section of CE$\nu $NS with respect to
the nuclear recoil energy $T$, for neutrinos with energy $E_{\nu }$
scattered off a target nucleus $(A,Z)$, can be written as 
\begin{equation}
\frac{d\sigma }{dT}(E_{\nu },T)\simeq \frac{G_{F}^{2}M}{\pi }Q_{W}^{2}\left(
1-\frac{MT}{2E_{\nu }^{2}}\right) F^{2}(q^{2})\,,
\label{eq:diff}
\end{equation}%
where $G_{F}$ is the Fermi constant, $M$ the nuclear mass, $T$ is nuclear recoil energy, and $Q_{W}^{2}$ is the weak nuclear charge 
\begin{equation}
Q_{W}^{2}=\left[ Zg_{p}^{V}+Ng_{n}^{V}\, \right]^{2} \,.  \label{eq:Qdef}
\end{equation}%
Here $Z$ is the proton number, 
$N$ the neutron number (tiny contributions from the sodium dopant of the detector can be ignored \cite{Akimov:2018vzs}), and the standard vector coupling constants are $%
g_{p}^{V}=1/2-2\sin ^{2}\theta _{W}$, $g_{n}^{V}=-1/2$. Finally,  $F(q^{2})$ is the nuclear form factor, we use the Klein-Nystrand parametrisation  \cite{Klein:1999qj,Engel:1991wq}:
\begin{equation}
F(q^{2})=\frac{4\pi \rho _{0}}{Aq^{3}}[\sin (qR_{A})-qR_{A}\cos
(qR_{A})]\left[ \frac{1}{1+a^{2}q^{2}}\right] . \label{F-bessel}
\end{equation}%
Here $q^{2}=2MT$ is the momentum transfer in the scattering of
neutrinos off the CsI nuclei, $\rho _{0}$ is the normalized nuclear
density, $R_{A}=1.2A^{1/3}\, \mathrm{fm}$ is the nuclear radius and $a=0.7\,%
\mathrm{fm}$ is the range of the Yukawa potential. 
Following Ref.~\cite{Akimov:2017ade} we will treat form factors entering the Cs and I cross-sections as the same. 

\begin{figure}[t]
\begin{center}
\includegraphics[width=5in]{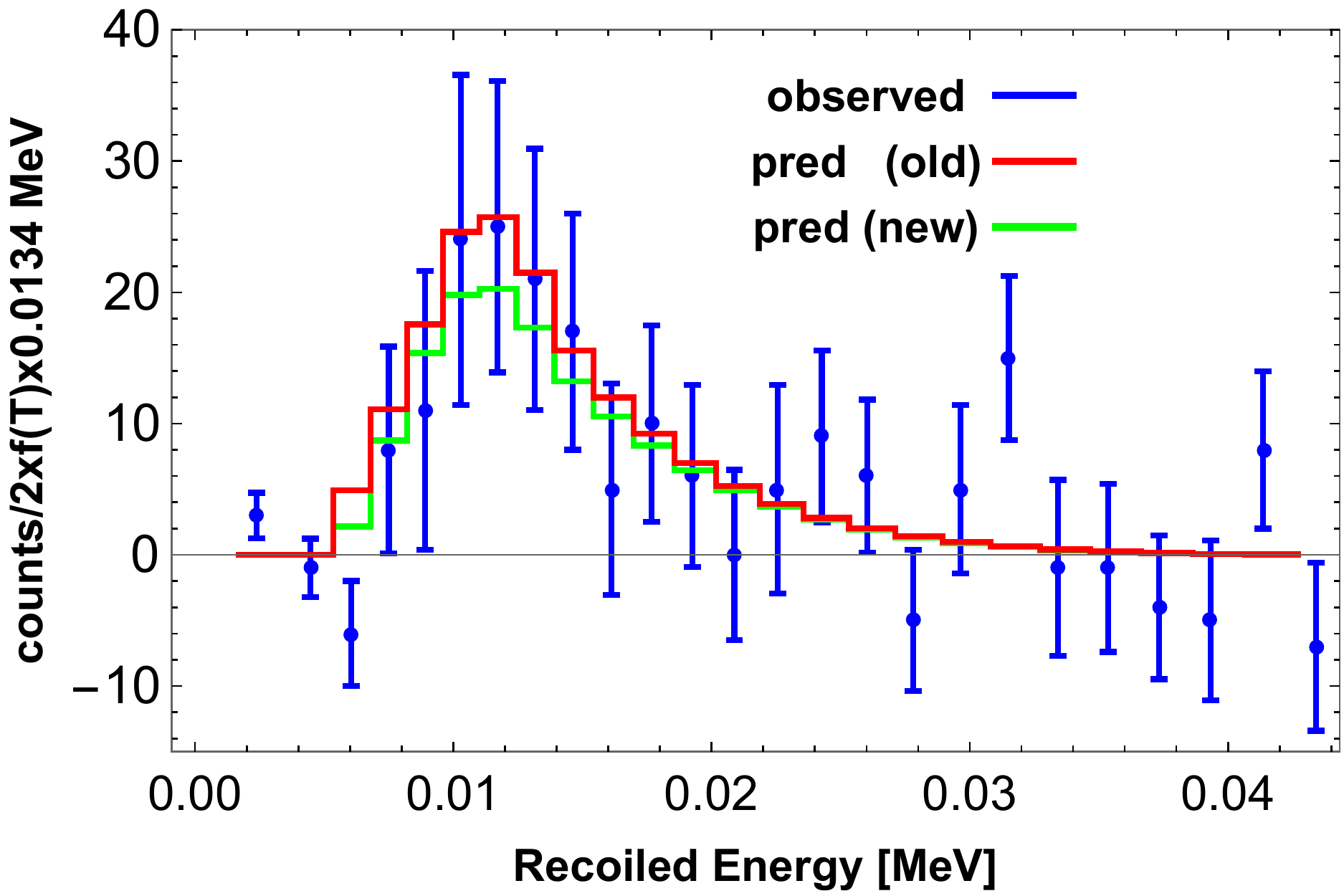}
\end{center}
\caption{The SM\ expected nuclear recoil  energy spectrum of CE$\protect \nu 
$NS for the COHERENT setup as function of the recoil energy. The points 
with the vertical error bars correspond to the COHERENT\ data. The
expected spectrum was obtained with the new quenching factor. The old spectrum (red) has been re-scaled horizontally for comparison with the new result (green).}
\label{Spectrum1}
\end{figure}

The differential event rate, after taking into account the detection
efficiency $\epsilon (T)$, taken from Fig.~S9 in Ref.~\cite{Akimov:2017ade}, of
COHERENT reads 
\begin{equation}
\frac{dN_{\nu _{\alpha }}}{dT}=t N\underset{E_{\nu }^{\min }}{\int
\limits^{E_{\nu }^{\max }}}dE_{\nu }\frac{d\sigma }{dT}(E_{\nu },T)\frac{%
d\phi _{\nu _{\alpha }}(E_{\nu })}{dE_{\nu }}\epsilon (T)\,,\label{eq:eventrt}
\end{equation}%
where $t=308.1$~days is the run time of the experiment, $N=\frac{2m_{%
\mathrm{det}}}{M_{\rm CsI}}N_{A}$ is the total number of target nucleons, $m_{%
\mathrm{det}}=14.57$ kg, $N_{A}$ is Avogadro's  number and $M_{\rm CsI}$ is the
molar mass of CsI.


In the first result of COHERENT \cite{Akimov:2017ade} the 
expected number of photo-electrons (p.e.) was 0.00117 p.e.\ ($T\!/\rm MeV$). 
The recent new measurement from Ref.\ \cite{Collar:2019ihs} improves this value and moreover gives its  energy dependence. We can use the 
following relation between the recoil energy and number of photo-electrons:
\begin{equation*}
N(p.e.) = f(T) \times 0.0134\  (T\!/{\rm MeV})\,,
\end{equation*}%
where $f(T)$ is the new quenching factor whose energy 
dependence is given in the left panel of Fig.\ 1 in 
Ref.\  \cite{Collar:2019ihs}. \footnote{We thank the authors of 
Ref.\ \cite{Collar:2019ihs} for providing us with the data.} 
For the acceptance function, we use Eq.\ (1) of Ref.\ \cite{Akimov:2018vzs}
as recommended there: 
\begin{equation}
\epsilon (T)=\frac{a_{1}}{1+\exp (-a_{2}(T-T_{0}))}\Theta (T)\,.
\label{eq:effic}
\end{equation}
Here $a_{1}=0.6655$, $a_{2}=494.2$ MeV$^{-1}$, 
$T_{0}=0.0092741$  MeV and the Heaviside function reads 
\begin{equation*}
\Theta (x)=\left \{ 
\begin{array}{c}
0\  \  \  \  \  \  \  \  \  \  \  \ x<5, \\ 
0.5\  \  \  \ 5\leq x<5, \\ 
1\  \  \  \  \  \  \  \  \  \  \  \  \  \  \ x\geq 6.%
\end{array}%
\right. 
\end{equation*}
All results in this paper will be derived by considering the
following $\chi ^{2}$-function: 
\begin{equation}
\chi ^{2}=\underset{i=4}{\overset{20}{\sum }}\frac{%
[N_{obs}^{i}-N_{exp}^{i}(1+\alpha )-B^{i}(1+\beta )]^{2}}{(\sigma ^{i})^{2}}%
+\left( \frac{\alpha }{\sigma _{\alpha }}\right) ^{2}+\left( \frac{\beta }{%
\sigma _{\beta }}\right) ^{2}\,.  \label{eq:chisq}
\end{equation}%
Here $N_{obs}^{i}$ is the observed event rate in the $i$-th energy bin, 
$N_{exp}^{i}$ is the expected event rate given in Eq.\ (\ref{eq:eventrt}) integrated over the recoiled energy corresponding to each flavor, and $B^{i}$  is the estimated 
background event number in the $i$-th energy bin extracted from Fig.\ S13 
of Ref.\  \cite{Akimov:2017ade}. The statistical uncertainty in the $i$-th energy bin is $\sigma ^{i}$, 
and $\alpha $, $\beta $ are the pull parameters related to the signal systematic 
uncertainty and the background rates. The corresponding uncertainties of 
the pull parameters are $\sigma _{\alpha }=0.28$ (previous value \cite{Akimov:2017ade}) 
$0.135$ (new value \cite{Collar:2019ihs}) and $\sigma _{\beta }=0.25$. 
We calculate $\sigma _{\alpha }$ by adding the flux 
uncertainty (10\%), neutron capture (5\%), acceptance (5\%), QF (25\%-old and 
5.1\%-new) in quadrature. The effect of the new quenching factor with the 
improved uncertainty on the recoiled energy spectrum is shown in Fig.\ \ref{Spectrum1} in red (old) and  green (new). 

Note that for simplicity we do not fit the prompt $\nu _{\mu }$ and the delayed $\nu_{e}$, 
$\bar{\nu}_{\mu }$ separately. In the plots that will be presented 
in what follows, our best-fit value is always indicated by a black dot. The total event rate we obtained with the above set by summing over all the energy bins are ~167 (previous) and ~139 (new) which are well within 1 sigma of the expected values of $173 \pm 48$ (old) and $138 \pm 19$ (new), respectively.

\section{\label{sec:EW}Constraints on electroweak physics of neutrinos}
\noindent 
In this section discuss the improved constraints on the Weinberg angle $\sin ^{2}\theta _{W}$, on 
parameters related to possible new  electromagnetic properties of neutrinos, and on the neutron rms charge radius. 

\subsection{Evaluation of $\sin^{2}\protect \theta _{W}$}

\begin{figure}[t]
\begin{center}
\includegraphics[width=7in,height=4.1in]{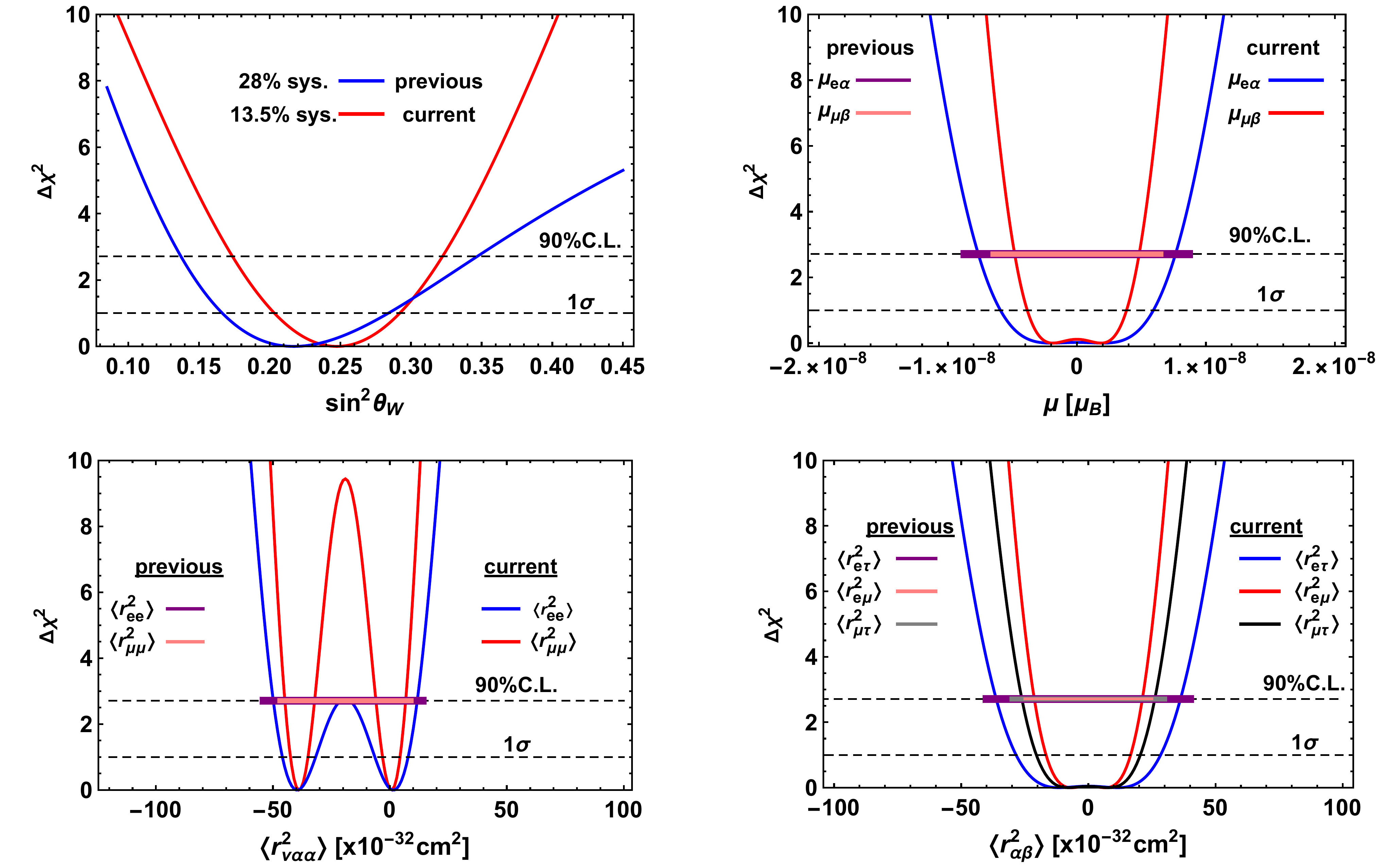}
\end{center}
\caption{1-dimensional $\Delta \protect \chi^{2}$-distributions of $\sin ^{2}\protect%
\theta _{W}$ (top-left), neutrino magnetic moments (top-right) and charge radii (lower left+right).   For $\sin ^{2}\protect \theta _{W}$ results with previous and current QF values are shown by separate curves, while for all other
cases the previous results are shown through horizontal lines at 90\% C.L. }
\label{1dim_all}
\end{figure}

Since the systematic effects are directly correlated with the electroweak physics 
parameters of CE$\nu $NS, any improvement in the quenching factor 
significantly affects for $\sin^{2}\theta _{W}$ its best-fit value and  uncertainty. The dependence on the Weinberg angle enters via $g_n^V$ in Eq.\  
(\ref{eq:Qdef}) in the differential cross section Eq.\  (\ref{eq:diff}). 
The $\Delta \chi ^{2}$-distributions of $\sin
^{2}\theta _{W}$ with old and new systematic uncertainties are 
displayed in the upper left plot of Fig.\ \ref{1dim_all}. It is evident from the figure that the central
value of $\sin^{2}\theta _{W}$ has significantly shifted towards a larger value. 
The new value from the COHERENT\ data with 
improved systematics is now
\begin{equation}
\sin ^{2}\theta _{W}=0.248\pm 0.045\, (1\sigma )\pm 0.074\, (90\%\,{ \rm C.L.}),
\end{equation}%
whereas the older value with 28\% systematic uncertainty is 
$\sin ^{2}\theta_{W}=0.217_{-0.051}^{+0.068} \,(1\sigma )_{-0.08}^{+0.13}  \,(90\%\ { \rm C.L.})$. 
The prediction of the modified $\overline{MS}$ 
renormalization scheme for sub-MeV momentum regime at low energy is  $\sin^{2}\theta _{W}=0.23867\pm
0.00016$ \cite{Erler:2004in} at 90\% C.L. 
The fact that COHERENT data with the original QF yields a value smaller than the SM prediction is consistent with Refs.\  \cite{Kosmas:2017tsq,Huang:2019ene,Cadeddu:2018izq}. The new fit-result has 
an about 20\% smaller error and is closer to the SM-prediction. The error is also much more Gaussian. 


\subsection{ Neutrino magnetic moments}
Magnetic moments appear in the general coupling of neutrinos to the electromagnetic field strength for Majorana ($M$) or Dirac ($D$) neutrinos 
\begin{equation}
{\cal L}^M = -\frac 14 \bar \nu_{\alpha L}^c \, \lambda_{\alpha \beta}^M \, \sigma_{\mu\nu} \, \nu_{\beta L} \, F^{\mu\nu}~\mbox{ or }~{\cal L}^D =-\frac 12 \bar \nu_{\alpha R} \, \lambda_{\alpha \beta}^D \, \sigma_{\mu\nu} \, \nu_{\beta L}\, F^{\mu\nu}\,.
\end{equation}
Here $\lambda^X = \mu^X - i \epsilon^X$, which is antisymmetric (hermitian) for Majorana (Dirac) neutrinos. Complex phases and $\epsilon^X$ are ignored here, see Ref.\ \cite{Miranda:2019wdy} 
for a general discussion. For Majorana neutrinos, in particular, there are only transition magnetic moments, 
$\mu^M_{\alpha\alpha}=0$. With unknown final state neutrino flavor no distinction between 
Dirac and Majorana neutrinos is possible. We assume here for definiteness Majorana neutrinos (and will drop the superscript $M$ from now on) and thus are sensitive to $\mu_{e \alpha}$ and $\mu_{\mu \beta}$ with $\alpha = \mu,\tau$ and $\beta = e,\tau$. 

\begin{figure}[t]
\begin{center}
\includegraphics[width=7.in]{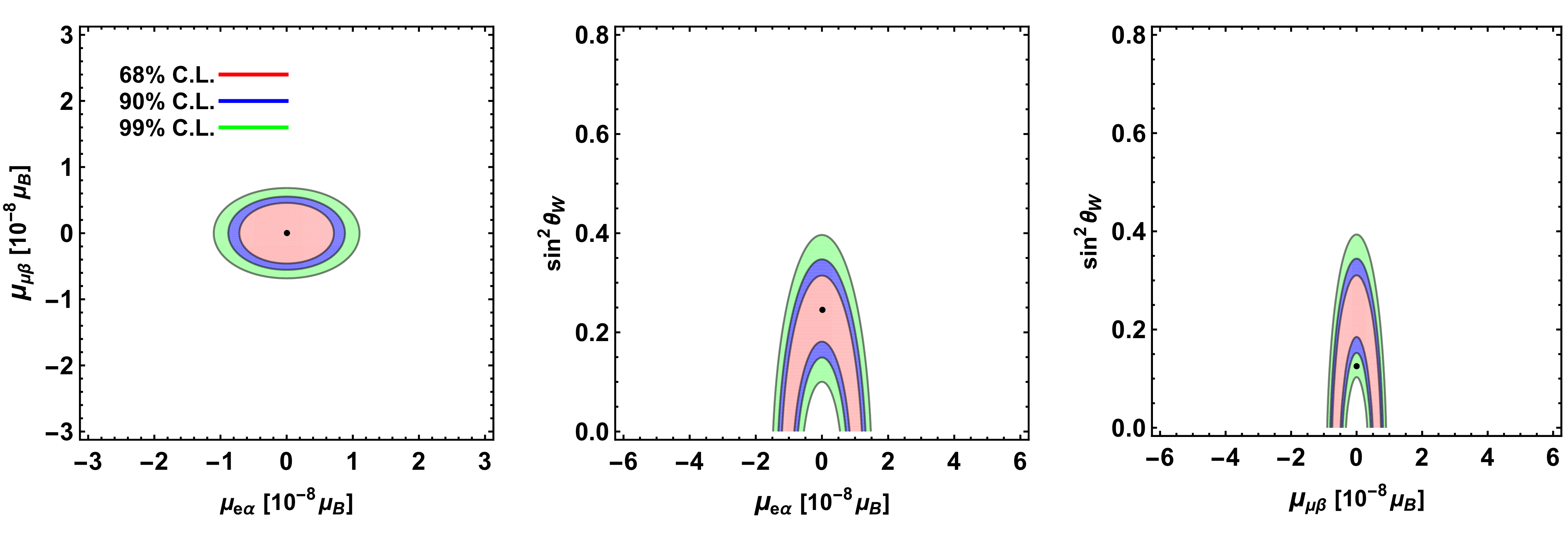}
\end{center}
\caption{2-dimensional $\Delta \protect \chi ^{2}$-contour plots for various
combinations of $\sin ^{2}\protect \theta _{W}$ and magnetic moments  
 with 68\%, 90\% and 99\% C.L.\ boundaries.}
\label{figmm2d}
\end{figure}

The contribution of a helicity-changing neutrino magnetic moment contribution adds to the helicity-conserving SM cross-section incoherently. Therefore we can make for the case of $\nu_e$ the  replacement $Q_{W}^{2}\rightarrow Q_{W}^{2}+Q_{mm, e}^{2}$, where 
$Q_W^2$ is given in Eq.\ (\ref{eq:Qdef}) and 
\begin{equation}
Q_{mm, e}^{2}=\left( \frac{\pi \alpha _{em}\, \mu _{e \alpha }\,Z}{2\sqrt{2}G_{F}m_{e}}\right)^{2}%
\text{ }\left( \frac{1}{T}-\frac{1}{E_{\nu }}+\frac{T}{4E_{\nu }^{2}}\right) \frac{8}{%
M\left( 1-\frac{MT}{2E_{\nu }^{2}}\right) }     \label{EDM}
\end{equation}
and analogously for $\nu_\mu$/$\bar\nu_\mu$. 
Here $\alpha _{em}$ is the fine-structure
constant, $m_{e}$ the electron mass and $\mu _{e\alpha }$ is the effective 
neutrino magnetic moment in units of Bohr magnetons $\mu _{B}$. The result of the fits is shown in Fig.\  \ref{1dim_all} (top-right) for one parameter at-a-time and in Fig.\ \ref{figmm2d} for two-parameter fitting. In the 1-dimensional plot, the previous constraints are 
shown for comparison at 90\% C.L.\ for each case. Improvement can be clearly seen for both parameters. The new constraints obtained from  one parameter at-a-time fitting at 90\% C.L.\ in units of $\mu _{B}$ are%
\begin{eqnarray}
-76\times 10^{-10} &<&\mu _{e \alpha}/\mu_B<76\times 10^{-10}\,, \\
-48\times 10^{-10} &<&\mu _{\mu\beta }/\mu_B<48\times 10^{-10}\,,
\end{eqnarray}%
while the previous constraints from our analysis are 
$-86\times 10^{-10} <\mu _{e\alpha}/\mu_B<86\times 10^{-10}$ and 
$-57\times 10^{-10} <\mu _{\mu\beta }/\mu_B<57\times 10^{-10}$, respectively. 
Improvement by 13\% and 20\% is found for $\mu_{e\alpha}$ and $\mu_{\mu\beta}$  when an improved QF is taken into account. 

\begin{figure}[h!]
\begin{center}
\includegraphics[width=5.5in,height=.7\textheight]{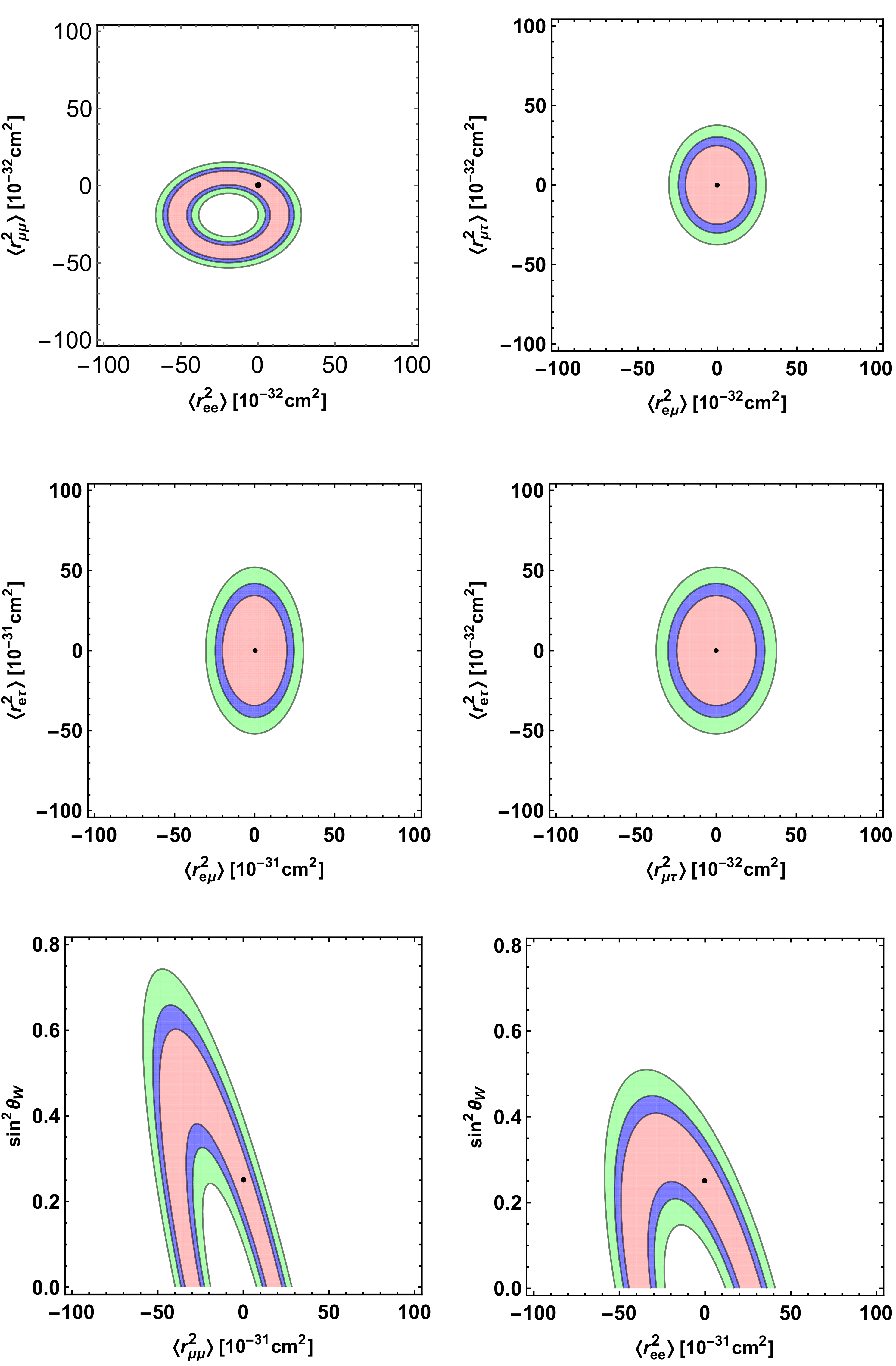}
\end{center}
\caption{2-dimensional $\Delta \protect \chi ^{2}$-contour plots for various 
combinations of $\sin ^{2}\protect \theta _{W}$ and neutrino charge radii 
with 68\%, 90\% and 99\% C.L.\  boundaries. }
\label{2d_cr}
\end{figure}

\subsection{Neutrino charge radii}

Massive neutrinos have an effective electromagnetic vertex $\bar \nu \Lambda_\mu \nu A^\mu$ with 
\cite{Giunti:2014ixa,Cadeddu:2018dux} 
\begin{equation*}
\Lambda _{\mu }(q)=\gamma _{\mu }F(q^{2}) \simeq \gamma 
_{\mu }q^{2}\frac{\langle r^{2}\rangle }{6}\,,
\end{equation*}
where $q$ is the momentum transfer and $F(q^{2}) $ is a form 
factor connected to the neutrino charge radius $\langle r^{2}\rangle $ via 
\begin{equation*}
\langle r^{2}\rangle =6\left. \frac{dF_{\nu }(q^{2})}{dq^{2}} \right|_{q^{2}=0} \,.
\end{equation*}
The expression in the SM \cite{Bernabeu:2000hf,Bernabeu:2002nw,Bernabeu:2002pd} is 
\begin{equation*}
\langle r_{\alpha\alpha }^{2}\rangle _{\rm SM}=-\frac{G_{F}}{2\sqrt{2}\pi }\left[
3-2\ln \left( \frac{m_{\alpha}^{2}}{m_{W}^{2}}\right) \right] ,
\end{equation*}
where $m_{\alpha}$ is the mass of the charged lepton associated to $\nu_{\alpha}$. Only diagonal charge radii $ \langle r_{\alpha\alpha }^{2}\rangle$ exist in the SM, while in general also transition charge radii $ \langle r_{\alpha\beta }^{2}\rangle$ are possible. The former add coherently to the 
 SM process, and we can take their effect into account by making for incoming neutrinos of flavor 
 $\alpha$ the replacement $g_{p}^{V}\rightarrow g_{p}^{V}+g_{em, \alpha}^{V},$ where $g_p^V$ is given above Eq.\ (\ref{eq:Qdef}) and 
\begin{equation}
g_{em,\alpha}^{V}=-\frac{\sqrt{2}\pi \alpha _{em}}{3G_{F}}\text{ }\langle
r_{\alpha \alpha }^{2}\rangle \,.  \label{crdig}
\end{equation}%
For the COHERENT\ setup, $%
\langle r_{ee}^{2}\rangle $ and $\langle r_{\mu \mu }^{2}\rangle $ are
relevant. The contribution of the flavor transition charge radii adds incoherently to the flavor-conserving SM process. Hence we can make for $\nu_e$ the replacement 
$Q_{W}^{2}\rightarrow Q_{W}^{2}+ Q_{em, e}^{2}$, where $Q_{em, e}^{2}$ is given 
by 
\begin{equation}
Q_{em,e}^{2}=\left( \frac{\sqrt{2}6\pi \alpha _{em}Z}{3G_{F}}\text{ }\langle
r_{e \alpha }^{2}\rangle \right) ^{2} \,  \label{crft}
\end{equation}
where $\alpha = \mu,\tau$.  While the neutrino flux at 
COHERENT\ includes $\nu_\mu$ and $\overline{\nu}_\mu$, since the transition charge radii of anti-neutrinos change only sign with respect to the ones for neutrinos \cite{Giunti:2014ixa}, 
only three flavor transition charge radii parameters are present: 
$\langle r_{e\mu }^{2}\rangle
,\  \langle r_{e\tau }^{2}\rangle $ and $\langle r_{\mu \tau }^{2}\rangle $. However, we have realized that this in principle is correct, but since the the weak neutral current couplings also change their signs from neutrinos to anti neutrinos under CP-transformation which leaves the overall sign of the term $g_{p}^{V}+g_{em, \alpha}^{V},$ unchanged. As a result, the sign changing for the neutrino charge radii for muon anti-neutrino has no effects and we get similar $\chi ^{2}$-distribution of $ \langle r_{\mu\mu }^{2}\rangle$ and $ \langle r_{\e\e }^{2}\rangle$ as shown in the lower left panel in fig. 2.

The results for one parameter at-a-time and  two parameter fitting 
are shown in Figs.\ \ref{1dim_all} (lower 2 panels) and \ref{2d_cr}, respectively. 
In Fig.\ \ref{1dim_all}, the results for 28\% systematic errors are shown using horizontal 
lines at 90\% C.L.\ for comparison. 
Improvement by (13-40)\% is found when an improved QF is taken into account. 
Our 90\% C.L.\ constraints on the neutrino charge radii, in units of cm$^{2}$, are
\begin{eqnarray}
-48\times 10^{-32} &<&\langle r_{ee}^{2} \rangle/{\rm cm^2} <12\times 10^{-32} \,,\nonumber \\
-44\times 10^{-32} &<&\langle r_{\mu \mu }^{2}\rangle/{\rm cm^2} <6\times 10^{-32}  \,,\nonumber \\
-8\times 10^{-32} &<&\langle r_{e\mu }^{2}\rangle/{\rm cm^2} <8\times 10^{-32} \,,\\
-18\times 10^{-32} &<&\langle r_{e\tau }^{2}\rangle/{\rm cm^2} <18\times 10^{-32} \,,\nonumber\\
-12\times 10^{-32} &<&\langle r_{\mu \tau }^{2}\rangle/{\rm cm^2} <12\times 10^{-32}\,.\nonumber
\end{eqnarray}


\subsection{Neutron Charge Radius and Cs weak nuclear charge}
Nuclear physics parameters can be tested by coherent scattering as well. 
We estimate here the neutron charge radius of CsI nuclei using the  improved QF following the prescription of Ref.\  \cite{Cadeddu:2017etk}. We use the form factor defined in Eq.\ (\ref{F-bessel}) both for protons and neutrons except that for neutrons we replace $R_{A}$ by 
\begin{equation*}
R_{A}=\sqrt{\frac{5}{3}(R_{n}^{2}-6a^{2})}\,,
\end{equation*}%
Here $R_{n}$ is the root-mean-square (rms) neutron charge radius. Notice that all results are  obtained in the approximation that the radii are the same for Cs and I. 

\begin{figure}[t]
\begin{center}
\includegraphics[width=3.5in]{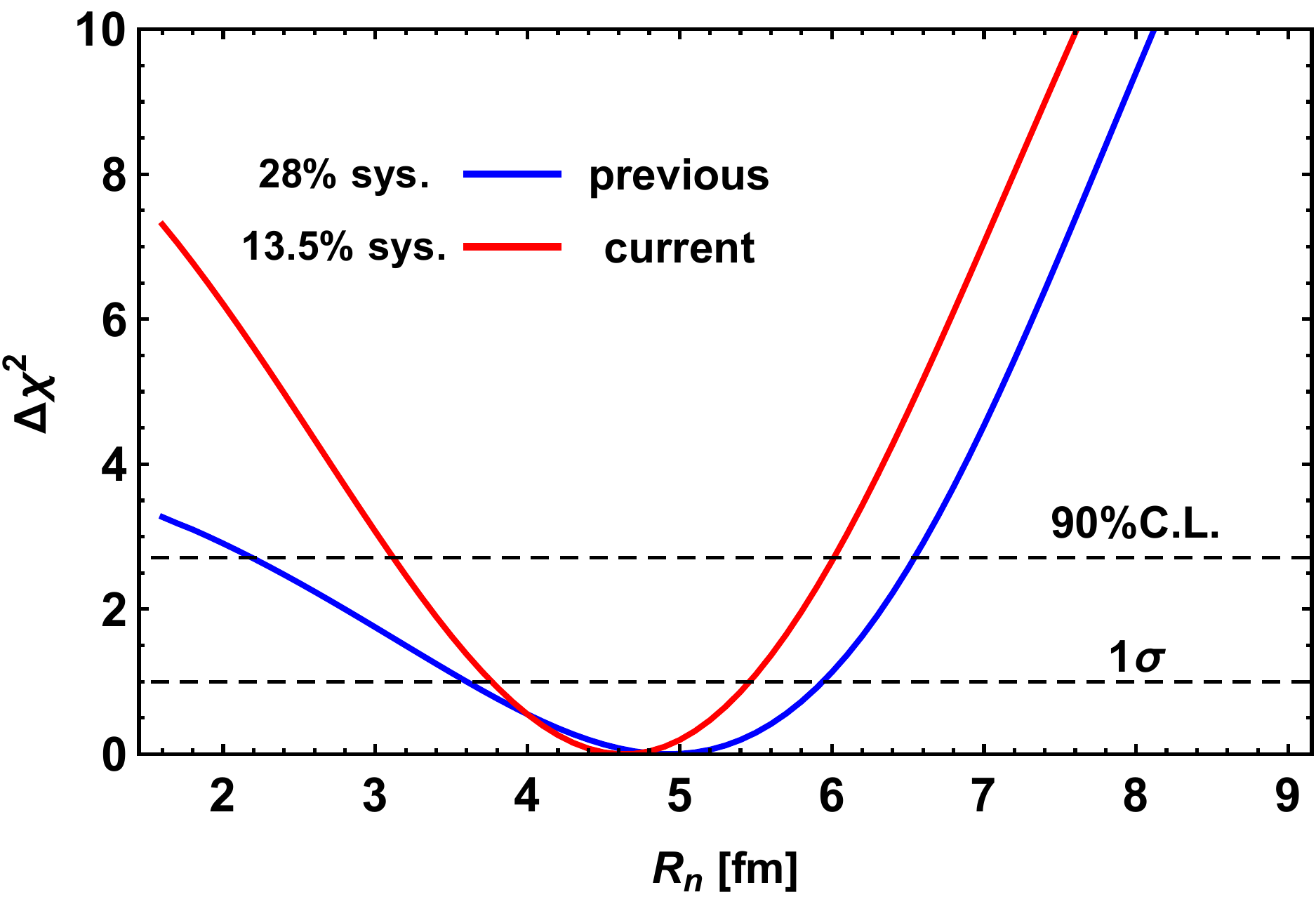}
\end{center}
\caption{1-dim $\Delta \protect \chi ^{2}$ for the neutron rms charge radius. }
\label{neutron_rad}
\end{figure}

We obtain the following best-fit values of the neutron charge radius of $^{133}$Cs and $^{127}$I 
\begin{eqnarray}
R_{n} &=&4.6_{-0.8}^{+0.9}\ {\rm fm}  \ (1\sigma )\    \  \ (\text{current}), \\
R_{n} &=&4.9_{-1.3}^{+1.1}\ {\rm fm} \ (1\sigma )\  \  \ (\text{previous}).
\end{eqnarray}
The $\Delta \chi ^{2}$-distribution of a one-parameter fit is shown in Fig.\ \ref{neutron_rad}. Notice that with the improved QF, there is 10\% improvement
in uncertainty, the distribution becomes more Gaussian, and the best-fit value is  shifted towards a relatively lower value.
Notice that the value obtained in Ref.\  \cite{Cadeddu:2017etk} was $R_{n} =5.5_{-1.1}^{+0.9}\ {\rm fm}$, which is consistent within $1\sigma$. 

We note at this point that Ref.\ \cite{Cadeddu:2019eta} appeared a few days after this work,  and 
that in particular the best-fit point of $R_n$ differs considerably. We find that this can be traced mainly 
to our use of 17 energy bins and the Klein-Nystrand form factor, compared to 12 bins and the Helm form factor in Ref.\ \cite{Cadeddu:2019eta}. Indeed, repeating our fit with 12 energy bins and the Helm form factor yields a best-fit value of $R_{n}  =4.9$ fm, compared to the value $R_{n} =5.0$  fm in Ref.\ \cite{Cadeddu:2019eta}.

The so-called neutron skin \cite{Horowitz:1999fk} is the difference between neutron and proton charge radii. The neutron skin influences among other things the equation of state of neutron stars \cite{Hagen:2015yea}. For the proton radius one takes the rather precisely known value $R_{p}=4.78$ fm  \cite{Fricke:1995zz} to obtain 
\begin{equation}
\Delta R_{np}=R_{n}-R_{p}\simeq -0.18_{-0.8}^{+0.9}\ {\rm fm.} \label{neutrino_Skin}
\end{equation}%
While the new best-fit value is now in better agreement 
with the predicted values of different models, which are in the regime 0.1 to 0.2 fm \cite{Horowitz:1999fk}, the uncertainty is still large. \\

Using the method described in Ref.\ \cite{Cadeddu:2018izq}, we can also calculate the 
electroweak nuclear charge of Caesium, whose value from our analysis is now

\begin{eqnarray}
Q_{W}^{Cs}=-72.2_{-1.2}^{+1.4}\  (1\sigma )\  \  \ \  \ (\text{current}), \label{Qvalue}\\
Q_{W}^{Cs}=-72.6_{-2.0}^{+1.9}\  (1\sigma )\  \  \ (\text{previous}).
\end{eqnarray}

\begin{figure}[t!]
\begin{center}
\includegraphics[width=5.5in]{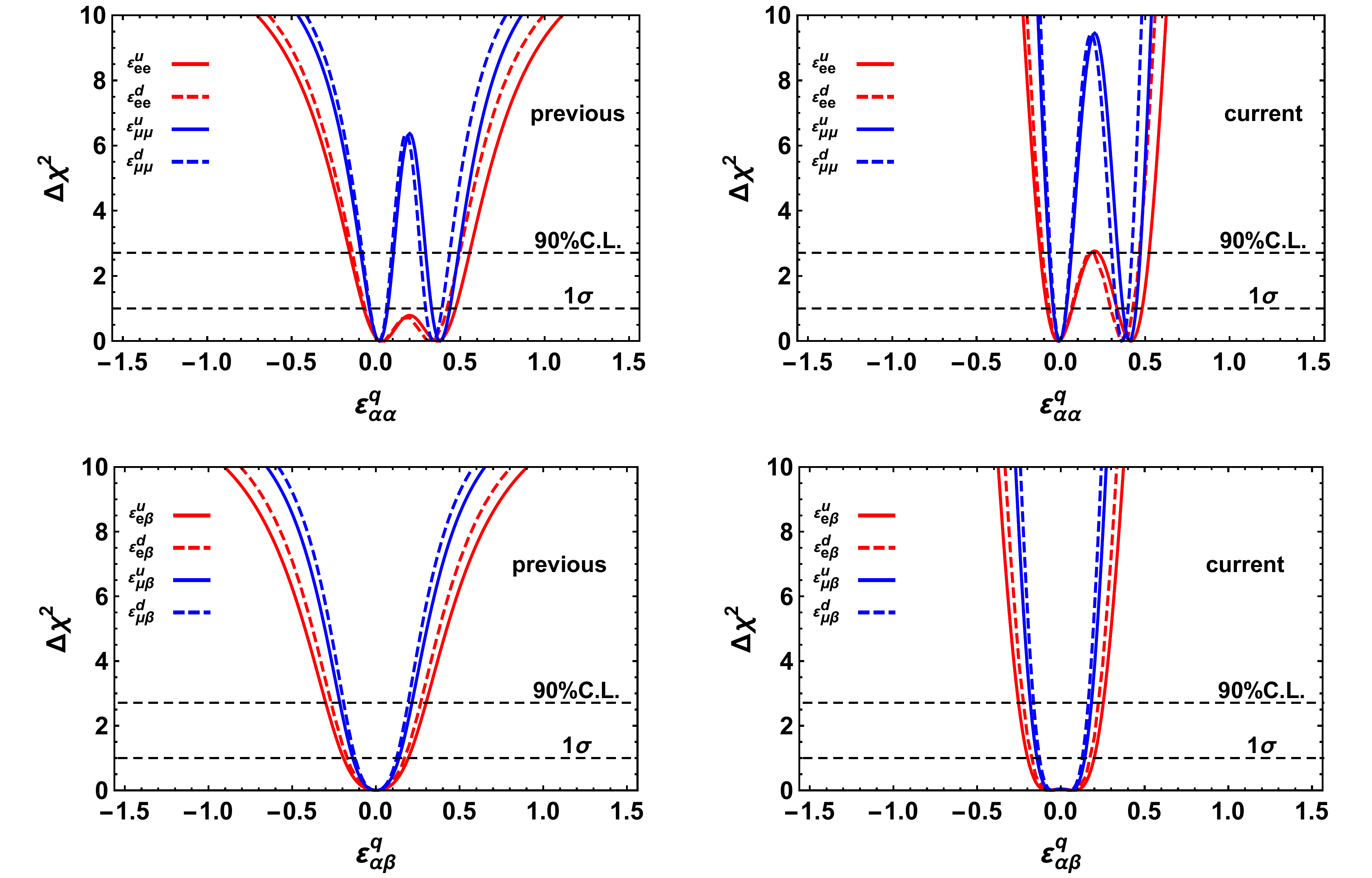}
\end{center}
\caption{1-dimensional $\Delta \protect \chi ^{2}$-distributions of neutrino-quark 
NSI. Note that for $\epsilon_{\mu\mu}$ the two degenerate solutions are now excluded at more 
than~3.5$\sigma$ with the improved QF.}
\label{fig_nsi_1d}
\end{figure}

\section{\label{sec:NSI}Neutrino Non-Standard Interactions}
\noindent
Non-Standard Interactions (NSI) of neutrinos are among the most often considered candidates for new neutrino physics \cite{Farzan:2017xzy,Dev:2019anc}. Motivated by their effects in neutrino oscillations one typically considers vector-like NSI in the form of dimension-6 operators:
\begin{equation}
-{\cal  L}_{NSI}    = \sqrt{8} \,G_F \, \epsilon_{\alpha \beta}^f 
\left(\bar \nu_{L \alpha} \gamma^\mu \nu_\beta\right) \left(\bar f \gamma_\mu f \right).
\end{equation}
The dimensionless parameters fulfill $\epsilon^f_{\alpha \beta} = \epsilon^{f\ast}_{\beta \alpha}$. 
For our purposes we need to consider $f = u,d$ and can distinguish flavor-diagonal (FD) and 
flavor-changing (FC) NSI. The FD case is treated in Eq.\ (\ref{eq:Qdef}) by making the replacement 
$Q_{W}^{2}\rightarrow Q_{W, \alpha \alpha }^{2}$, while for the FC case we use 
$Q_{W}^{2}\rightarrow Q_{W}^{2}+Q_{W, \alpha \beta }^{2}$: 
\begin{eqnarray}
Q_{W, \alpha \alpha }^{2} &=& \left[Z(g_{p}^{V}+2\varepsilon _{\alpha \alpha 
}^{u}+\varepsilon _{\alpha \alpha }^{d})+N(g_{n}^{V}+2\varepsilon _{\alpha
\alpha }^{d}+\varepsilon _{\alpha \alpha }^{u})\,\right]^{2} , \\
Q_{W, \alpha \beta }^{2} &=&\underset{\beta \neq \alpha }{\sum }%
\left|Z(2\varepsilon _{\alpha \beta }^{u}+\varepsilon _{\alpha \beta
}^{d})+N(2\varepsilon _{\alpha \beta }^{d}+\varepsilon _{\alpha \beta
}^{u})\right|^{2}.
\end{eqnarray}
Ignoring phases we explicitly write out the coupling factors for $\alpha =e$ and $\mu $: 
\begin{eqnarray*}
Q_{W, ee}^{2} &=&\left[Z(g_{p}^{V}+2\varepsilon _{ee}^{u}+\varepsilon
_{ee}^{d})+N(g_{n}^{V}+2\varepsilon _{ee}^{d}+\varepsilon
_{ee}^{u})\,\right]^{2}, \\
Q_{W, \mu \mu }^{2} &=&\left[Z(g_{p}^{V}+2\varepsilon _{\mu \mu }^{u}+\varepsilon
_{\mu \mu }^{d})+N(g_{n}^{V}+2\varepsilon _{\mu \mu }^{d}+\varepsilon
_{\mu \mu }^{u})\,\right]^{2}, \\
Q_{W, e\beta }^{2} &=&\left[Z(2\varepsilon _{e\beta }^{u}+\varepsilon _{e\beta
}^{d})+N(2\varepsilon _{e\beta }^{d}+\varepsilon _{e\beta }^{u})\,\right]^{2} 
\text{ (}\beta =\mu ,\tau \text{)} \,,\\
Q_{W, \mu \beta }^{2} &=&\left[Z(2\varepsilon _{\mu \beta }^{u}+\varepsilon _{\mu
\beta }^{d})+N(2\varepsilon _{\mu \beta }^{d}+\varepsilon _{\mu \beta
}^{u})\,\right]^{2}  \text{(}\beta =\mu ,\tau \text{)} \,, 
\end{eqnarray*}
where summation over $\beta $ in the last two lines is understood.

\begin{figure}[t!]
\begin{center}
\includegraphics[width=5.5in]{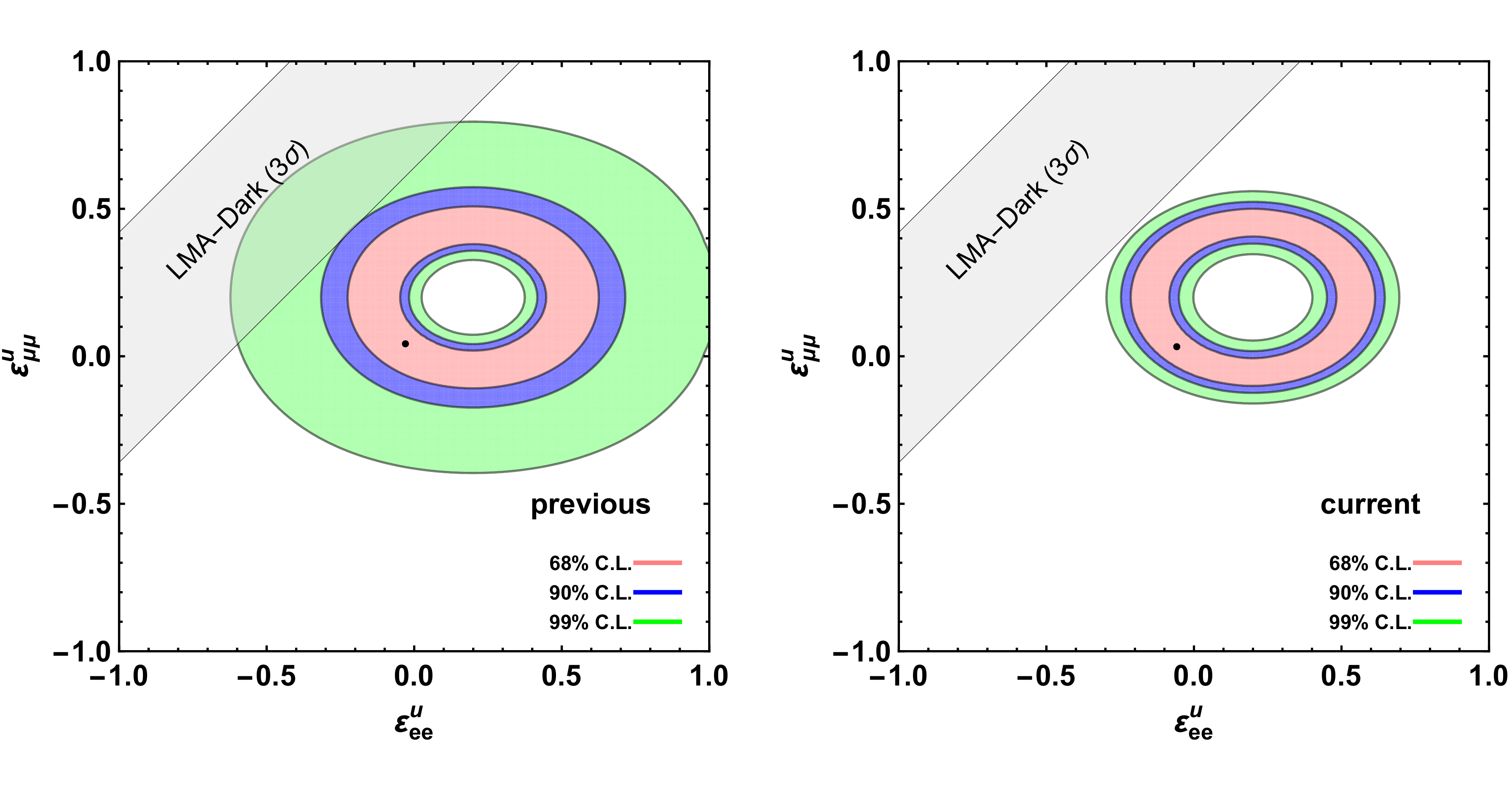}
\end{center}
\caption{2-dimensional $\Delta \protect \chi ^{2}$-contour regions of NSI parameters relevant for 
LMA-Dark, whose parameters have been overlaid on the COHERENT results. 
}
\label{fig_nsi_2d}
\end{figure}

The results of one parameter at-a-time fits for the NSI parameters  are shown in Fig.\ \ref{fig_nsi_1d}.  The bounds  at 90\% C.L.\ are 
\begin{eqnarray*}
\text{FD} &\text{:}&-0.12<\varepsilon _{ee}^{u}<0.52,\  \ -0.11<\varepsilon
_{ee}^{d}<0.47,\ \ -0.07<\varepsilon _{\mu \mu }^{u}<0.47,\  \ 
-0.06<\varepsilon _{\mu \mu }^{d}<0.42 \,, \\
\text{FC} &\text{:}&-0.25<\varepsilon _{e\beta }^{u}<0.25,\  \
-0.23<\varepsilon _{e\beta }^{d}<0.23,\ \ -0.18<\varepsilon _{\mu \beta
}^{u}<0.18,\  \  -0.16<\varepsilon _{\mu \beta }^{d}<0.16\,.
\end{eqnarray*}%
Of particular interest is a set of parameter values that would allow the LMA-Dark solution 
\cite{Miranda:2004nb} with a solar neutrino mixing angle $\sin^2 \theta_{12} > \pi/4$. 
It would correspond to large and negative $\epsilon_{ee} - \epsilon_{\mu\mu} = -{\cal O}(1)$. One such case is displayed in Fig.\ \ref{fig_nsi_2d}, compared to our fit results.  
A full analysis to quantify the degree with which LMA-Dark is ruled out would require fitting COHERENT  data together with neutrino oscillation experiments as done in Ref.\ \cite{Coloma:2017ncl}, and as shown  there the LMA-Dark solution caused by effective operators is ruled out by COHERENT (with the previous QF) at~3$\sigma$. 
Here we simply take the LMA-Dark allowed parameter values and compare with our fit. 
One can see from Fig.\ \ref{fig_nsi_2d} that the boundaries from our two-parameter fitting exclude the LMA-Dark solution at about 90\% C.L.\ (at 2.1 $\sigma$) for the  previous data (left figure), while for the new QF the exclusion occurs at more than 99\% C.L.\ (3.1 $\sigma$).

\section{\label{sec:bosons}New Neutral Currents from Vector and Scalar Mediators}
\noindent 
New neutral vector and scalar mediators may couple to neutrinos and quarks, thereby generating 
new neutral currents. We can write \cite{Cerdeno:2016sfi}
\begin{eqnarray}
{\cal L}_{vec}&=& g_{Z'} \left(\bar \nu_L \gamma^\mu \nu_L + \bar q \gamma^\mu q\right) Z'_\mu \,, \\
{\cal L}_{sca}&=& g_{\phi} \left(\bar \nu_R \nu_L + \bar q q\right) \phi + h.c. 
 \label{lagvecsca}
\end{eqnarray}%
For the vector case we restrict ourselves to the simplest scenario of coupling only to the 
left-handed SM neutrinos. We also assume all couplings to be universal. Apart from the couplings 
$g_{Z', \phi}$ we also have the masses $M_{Z', \phi}$ as new parameters.

\begin{figure}[t!]
\begin{center}
\includegraphics[width=5.5in]{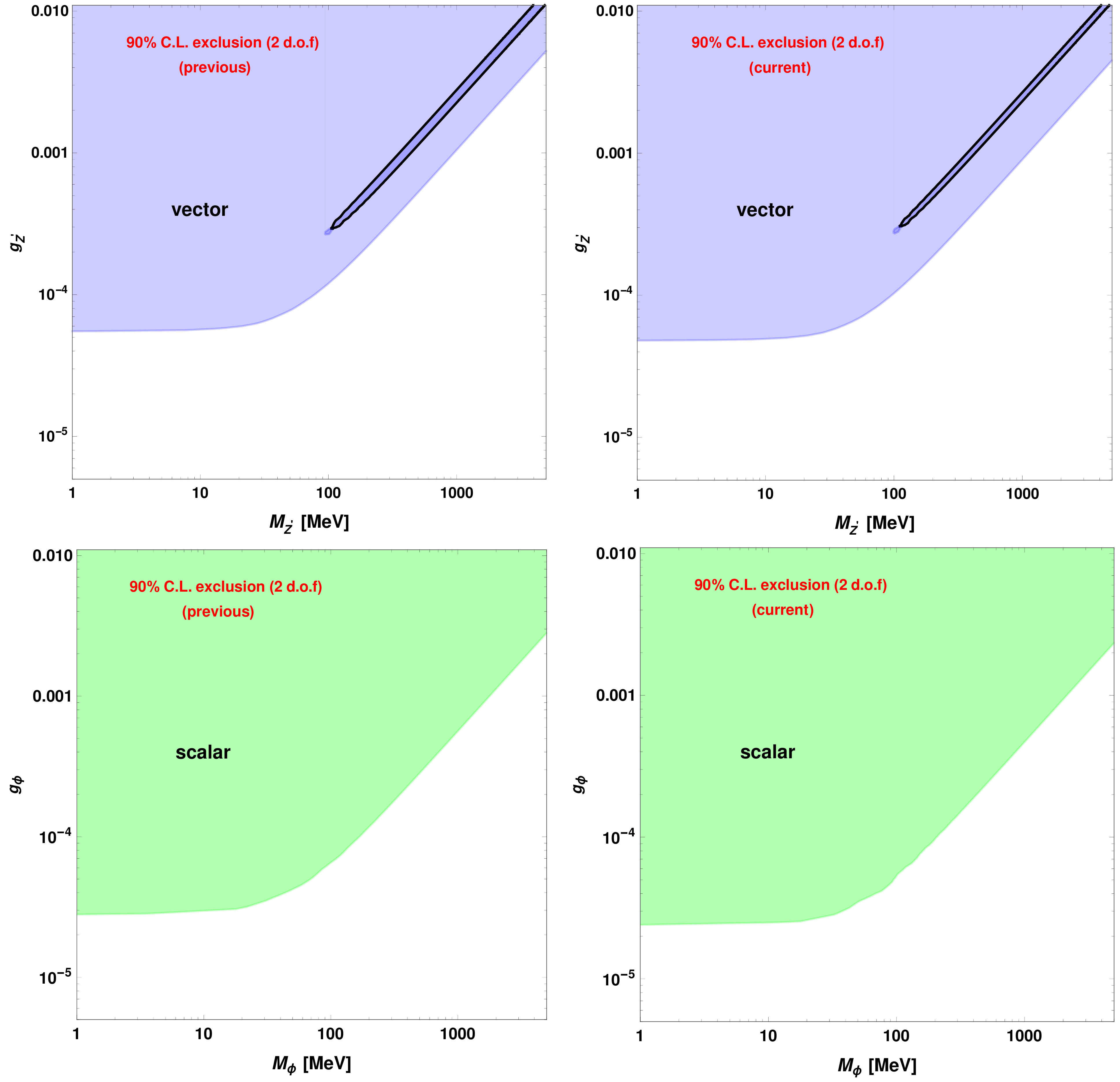}
\end{center}
\caption{2-dimensional $\Delta \protect \chi ^{2}$-contour plots at 90\% C.L.\ for 
 vector and scalar mediator masses and couplings with 28\% 
(left-panels) and 13.5\% systematic errors (right-panel) at 90\% C.L. }
\label{fig_mad}
\end{figure}

We can take new vector bosons into account by replacing the SM couplings constants in 
Eq.\ (\ref{eq:Qdef}) as ($g_{p}^{V}, g_{n}^{V}) \rightarrow  (g_{p}^{V}+ \widetilde{g}^{V},  g_{n}^{V}+\widetilde{g}^{V})$, where 
\begin{equation}
\widetilde{g}^{V}=\frac{3g_{V}^{2}}{2\sqrt{2}G_{F}(q^{2}+M_{Z^{^{\prime
}}}^{2})} \,.
\label{weak_vec}
\end{equation}%
The scalar contribution, in turn, is added to the cross-section incoherently via the 
replacement $Q_{W}^{2}\rightarrow Q_{W}^{2}+ Q_{sca}^{2}$, where 
\begin{equation}
Q_{sca}^{2}=\left( \frac{g_{\phi }^{2}(14N+15.1Z)}{2\sqrt{2}G_{f}E_{\nu
}(q^{2}+M_{\phi }^{2})}\right) ^{2}\frac{2MT}{\left( 1-\frac{MT}{2E_{\nu
}^{2}}\right) } \,.
\label{weak_sca}
\end{equation}
We take here vector and scalar weak charges in Eq.\ (\ref{weak_vec}) and (\ref{weak_sca}) from  calculations given in Ref.\ \cite{Cerdeno:2016sfi}. For COHERENT with 28\% (previous) 
and 13.5\% (current) 
systematic uncertainties, we show the results both for the vector and scalar
masses versus the coupling constants in Fig.\  \ref{fig_mad}. \\

Improvement can be seen from the plots, and in the vector case the degeneracy region \cite{Liao:2017uzy} (when $3g_V^2/M_{Z'}^2 = -G_F 4\sqrt{2} (Z g_p^V + N g_n^V)/(Z+N)$) shrinks down further, but does of course not wash out completely.

\section{\label{sec:concl}Conclusions}
\noindent 
Coherent elastic neutrino-nucleus scattering is an exciting new window to neutrino and neutral current physics. 
We investigated the effect of an improved quenching factor knowledge applied to COHERENT's measurement of the process. Several Standard Model and beyond the Standard Model parameters were considered. Improvement is found for all parameters, demonstrating again that the process is a powerful new handle to test many scenarios. Future measurements with higher statistics will further cement this.\\[1cm]

\noindent
\underline{\it \bf Note added:} When this paper was finalized, Ref.\ \cite{Papoulias:2019txv} appeared, which also uses the new QF measurement to probe several parameters in and beyond the Standard Model, although with a single-bin analysis while we do the full spectral analysis. Their results results agree with the relevant parts of this work except a few cases where different approaches were adopted.

\begin{acknowledgments}
\noindent 
We thank Grayson Rich and Juan Collar for useful discussions and
sharing the COHERENT data. WR was supported by the DFG with grant RO 2516/7-1 in the
Heisenberg program. AK is supported by the Alexander von Humboldt
foundation. AK is thankful to Evgeny Akhmedov, Carlo Giunti and Douglas McKay for useful discussions. 
\end{acknowledgments}

\bibliographystyle{apsrev4-1}
\bibliography{refs}

\end{document}